\documentclass{article}

     \PassOptionsToPackage{numbers, compress}{natbib}

 \usepackage[final]{neurips_2019_ml4ps}

\usepackage[utf8]{inputenc} 
\usepackage[T1]{fontenc}    
\usepackage{hyperref}       
\usepackage{url}            
\usepackage{booktabs}       
\usepackage{amsfonts}       
\usepackage{nicefrac}       
\usepackage{microtype}      

\usepackage{amssymb,braket,amsmath,mathrsfs,amsthm}
\usepackage{graphicx}

\title{Optimal Real-Space Renormalization-Group Transformations with Artificial Neural Networks}

\author{%
  Jui-Hui Chung \\
  Department of Physics\\
  National Taiwan University\\
  Taipei 10607, Taiwan \\
  \And
  Ying-Jer Kao \\
  Department of Physics\\
  National Taiwan University\\
  Taipei 10607, Taiwan \\
  \texttt{yjkao@phys.ntu.edu.tw} \\
}

\begin{document}

\maketitle

\begin{abstract}
We introduce a general method for optimizing real-space renormalization-group transformations to study the critical properties of a classical system. 
The scheme is based on minimizing the Kullback-Leibler divergence between the distribution of the system  and the normalized normalizing factor of the transformation parametrized by a restricted Boltzmann machine.
We compute the thermal critical exponent of the two-dimensional Ising model using the trained optimal projector and obtain a very accurate thermal critical exponent  $y_t=1.0001(11)$ after the first step of the transformation.
\end{abstract}

\section{Introduction}

Deep learning (DL)~\cite{lecun2015deep} has yielded impressive results in difficult machine learning tasks and various fields of physics~\cite{carleo2017solving,Carrasquilla:2017iza}. 
Despite its success, theoretical understanding of the reason behind the surprising effectiveness of DL is still lacking. 
Although a connection between the  renormalization group (RG) and the deep neural networks has been established~\cite{mehta2014exact},
it is desirable to construct a scheme to enable learning in the RG procedure in order to extract useful information, such as the critical exponents.

Monte Carlo renormalization group (MCRG)~\cite{swendsen1979monte}  is a promising computational scheme for the real-space renormalization group (RSRG). 
The major source of systematic errors in the MCRG calculations is the lack of convergence due to  slow approach to the fixed point. 
Attempts have been made to introduce variational parameters into the RG transformations with an optimal criterion to bring the fixed point closer to the nearest-neighbor model~\cite{swendsen1984optimization}.
The interpretation of why this proposal works, however, remains controversial~\cite{fisher1986location}.

In this paper, we propose a general method for optimizing RSRG transformation through divergence minimization using neural network. 
In our approach, the weight factor is parametrized with a restricted Boltzmann machine and the parameters are chosen to minimize the Kullback-Leibler (KL) divergence between the distribution of the system and the normalized normalizing factor of the weight factor. 

\section{Related Work}
Metha and Schwab~\cite{mehta2014exact}  established an exact mapping between the variational renormalization group~\cite{kadanoff1976variational} and deep neural networks based on RBM. 
%
%
They then applied  deep learning techniques to numerically coarse-grain the two-dimensional nearest-neighbor Ising model on a square lattice, and 
showed the scheme corresponds to implementing a coarse-graining scheme similar to block spin renormalization~\cite{kadanoff1976variational}.
Therefore, they suggested there exists a connection between  RG schemes and deep learning algorithms that minimize the Kullback-Leibler (KL) divergence.

On the other hand, Koch-Janusz and Ringel~\cite{koch2018mutual} claimed that  training RBMs by minimizing the KL divergence  does not perform RG. 
Instead, they proposed an information-theoretic characterization scheme that maximizes the  real-space mutual information (RSMI),  which is capable of generating samples of the coarse-grained system. 
After several RG transformations, they were able to extract the correlation length critical exponent $\nu=1.0\pm0.15\ ( y_t=1/\nu =1.0 \pm 0.15 )$.
We note, however, although the RSMI algorithm was used to generate a sequence of configurations, they did not use the standard MCRG technique~\cite{swendsen1979monte} to extract quantitative results from these configurations.

In our work, we demonstrate that applying divergence minimization in training RBM can generate an optimal RG transformation that filters out long-range coupling parameters. 

\section{Optimal Criterion}

A generic lattice-model Hamiltonian has the form
\begin{equation}
	{H}(\sigma)=\sum_\alpha K_\alpha S_\alpha(\sigma),
\end{equation}
where the interactions $S_\alpha$ are combinations of the original spins $\sigma$ and the $K_\alpha$ are the corresponding coupling constants. 
Consider a general RG transformation \cite{niemeijer1976renormalization}
\begin{equation}
	e^{{H'}(\mu)} = \sum_\sigma P(\mu|\sigma) e^{{H}(\sigma)}, \label{eq:generic_rg_transform}
\end{equation}
with a parametrized weight factor of the form
\begin{equation}
P(\mu |\sigma) = \frac{1}{Y(\sigma)} e^{\sum_{ij}W_{ij}\sigma_i\mu_j}, \label{eq:proj}
\end{equation}
where the normalization factor is
\begin{equation}
	Y(\sigma) = \prod_j  2\cosh \left[ \sum_{i}W_{ij}\sigma_i \right]. \label{normalizing}
\end{equation}
Here $\mu=\{+1,-1\}$ are the renormalized spins of the renormalized Hamiltonian ${H'(\mu)=\sum_\alpha K'_\alpha S_\alpha(\mu)}$ with renormalized coupling parameters $K'_\alpha$. 
And $W_{ij}$ are the variational paramters whose optimal criterion for choosing them is to be discussed. 
In particular, if $W_{ij}$ are infinite in a local block of spins and are zero otherwise, then we have the usual majority rule transformation \cite{swendsen1982}.

To motivate the optimal crtierion for choosing the variational parameters, we use a heuristic argument of why deep learning \cite{lecun2015deep} works so well. 
In deep learning, it is believed that an RBM machine \cite{goodfellow2016deep,bengio2007greedy}
\begin{equation}
	Q(v,h) = \frac{1}{Z} e^{\sum_{ij}W_{ij}v_i h_j}, \label{eq:rbm}
\end{equation}
parametrized by weights $W_{ij}$ with hidden variables $h_j$ and visible variables $v_i$, works to extract empirical feature distribution $\hat{p}'(h)$ from the empirical distribution $\hat{p}(v)$ through
\begin{equation}
	\hat{p}'(h) = \sum_{\sigma} Q(h|v)\hat{p}(v), \label{eq:RBM_extract_empirical}
\end{equation}
where $Q(h|v)\equiv Q(v,h)/\sum_v Q(v,h)$ is the conditional distribution of the hidden variables given the values of the visible variables. 
In this scheme, the RBM's parameters are chosen to minimize the KL divergence between the empirical distribution $\hat{p}(v)$ and the marginal distribution $\sum_{h}Q(v,h)$
\begin{equation}
	D\left(\hat{p}(v) \middle\| \sum_{h}Q(v,h) \right),
\end{equation}
where the KL divergence is defined as $D(p\| q) \equiv \sum_{\sigma} p(\sigma) \log (p(\sigma)/q(\sigma))$ for two discrete distribution $p(\sigma)$ and $q(\sigma)$. 
Hidden layers of an RBM are supposed to extract meaningful features from the data \cite{krizhevsky2009learning}.

Based on the similarity of equations~(\ref{eq:RBM_extract_empirical}) and~(\ref{eq:generic_rg_transform}), we identify the conditional distribution $Q(h|v)$ with our parametrized weight factor ${P}(\mu|\sigma)$ and associate the hidden and visible variables of an RBM with the renormalized and original spins respectively. 
The ``marginal distribution'' of the weight factor is identified as the normalized normalizing factor $Y(\sigma)/\sum_\sigma Y(\sigma).$
In analogy to the paramter-choosing scheme of an RBM machine, we propose an optimal criterion for choosing the parameter of a weight factor: minimize the KL divergence between the distribution of the system and the normalized normalizing factor of the weight factor
\begin{equation}
D\left(\frac{1}{Z}e^{{H}(\sigma)}  \middle\| \frac{Y(\sigma)}{\sum_\sigma Y(\sigma)} \right). \label{eq:kl_norm}
\end{equation}
The optimization problem is solved in the stochastic setting where we use machine learning and contrastive divergence algorithms.

\begin{figure}
\centering
\includegraphics{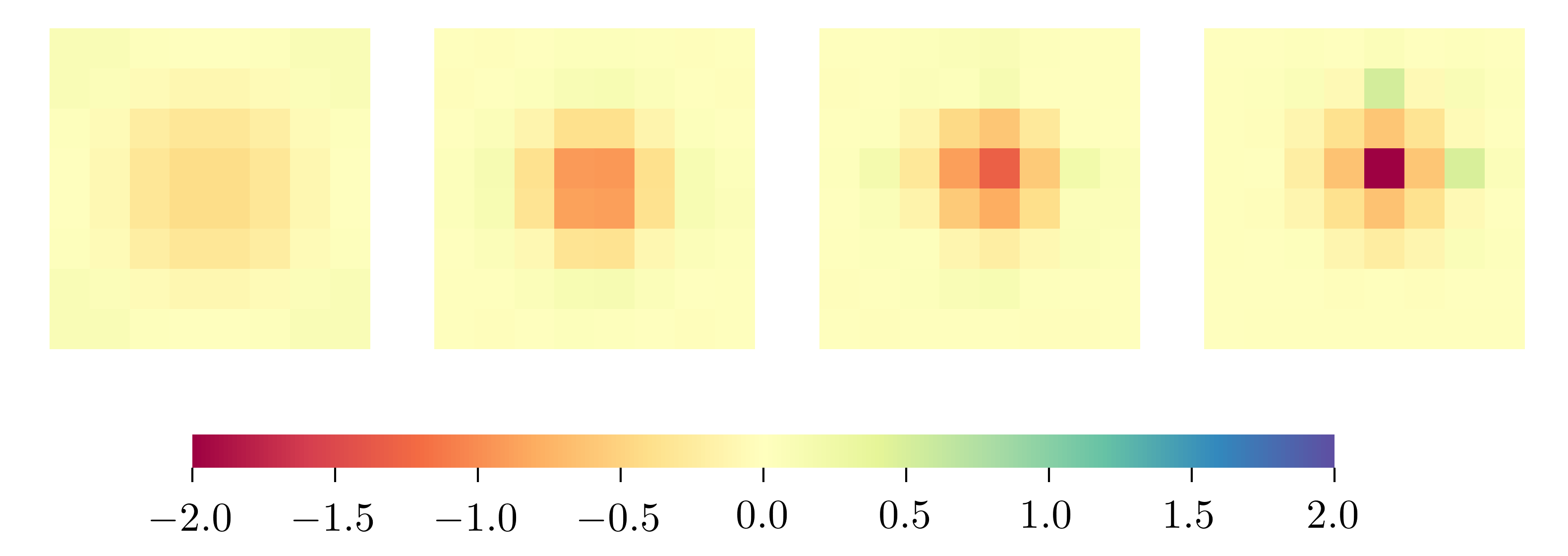}
\caption{\label{fig:filter}Machine representation of the optimal weight factor for a 2D Ising model. We show the feature maps for the $32\times 32$ Ising model at the critical temperature. From left to right we show the development of the feature map as the training progress (feature maps shown are at the training epoch of $1,3,5$ and $50$). The feature maps act as effective filters on the spin configurations, capturing the most important correlations.}
\end{figure}

\section{Results}
To validate our scheme, we consider the problem of finding the thermal critical exponent of the Ising model. The Hamiltonian is
\begin{equation}
{H}(\sigma) = K_{\text{nn}} S_{\text{nn}} = K_{\text{nn}}\sum_{\braket{ij}} \sigma_i \sigma_j,
\end{equation}
where $\sigma_i=+1$ or $-1$ and $K_{\text{nn}}$ is the nearest-neighbor coupling. In the following we consider a two dimensional lattice of size $32\times 32$ with periodic boundary conditions.

We prepare a data set with $10^4$ binary spin configurations sampled at the critical temperature. We update the parameters with contrastive divergence CD$_3$ and with an adaptive variant of stochastic gradient descent method called ADAM \cite{kingma2014adam}. The learning rate is initially set to $\eta=10^{-3}$ and decays during learning. A square root decay is applied to the initial learning rate to reach a finial value of $10^{-4}$ at the 25th epoch and the rate stays constant for the rest of 25 epochs. The minibatch contains 10 samples and the parameters are initialized uniformly around zero. Seven coupling terms were chosen according to Ref.~\cite{swendsen1982} for the MCRG analysis.

\begin{figure}
\centering
\includegraphics{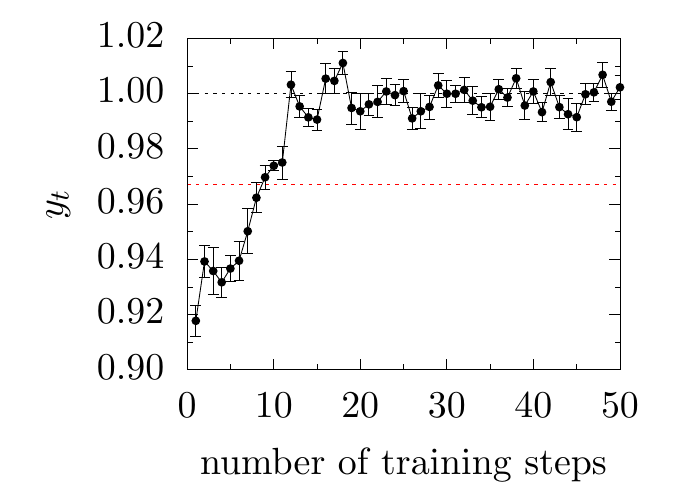}
\caption{\label{fig:iter}Thermal critical exponents $y_t$ over training steps on a $32\times 32$ lattice for the first step of renormalization. The red (lower) dotted line is the corresponding result $y_t=0.967(3)$ for the majority-rule transformation~\cite{swendsen1982} and the black (upper) dotted line is the exact value of $y_t=1$.}
\end{figure}

In Fig. \ref{fig:filter} we show the optimal machine structure of the weight factor learned on the Ising model with a filter size of $8\times 8$ and with imposed translational symmetries. We find that the filter learns localized feature which is in agreement with the conventional wisdom that renormalized spins and original spins which are close to one another should be coupled more strongly than others \cite{hilhorst1979exact}. For example, an extreme case of a localized machine is that of a $2\times 2$ filter with infinite weights, which is equivalent to the typical majority-rule transformation \cite{swendsen1982}. The behavior of our learned machine, on the other hand, also shows non-local interactions.

In Fig. \ref{fig:iter} we use our optimal machine to calculate the thermal critical exponent for the $32\times 32$ Ising model as a function of the training steps for the first step of renormalization transformation. The data used to compute the thermal exponenent is different from that used in training and consists of $5\times 10^4$ samples. The exponent converges to the exact value  (black dashed horizontal line) upon increasing training steps. The most striking result is that although the weight factors are \textit{learned} without any \textit{prior} knowledge of the system, they are able to generate a renormalization transformation such that the exponent approaches very close to the exact value after the first step of  the transformation. 

\begin{table}
\caption{MCRG estimates for the thermal critical exponents of the 2D Ising model from a simulation on a $32\times 32$ lattice using our optimal weight factor as compared to the majority-rule transformation. $N_r$ is the number of RG steps and $N_c$ is the number of couplings included in the analysis.}
\label{tab:expo}
\centering
\begin{tabular}{cccccccc}
\toprule
$N_r$ & $N_c$ & optimal & majority & $N_r$ & $N_c$ & optimal & majority \\[0.5ex]
\midrule
1 & 1 &  0.9217(09) & 0.904(1) & 2 & 1 &  0.9281(08) & 0.953(2) \\
  & 2 &  0.9910(08) & 0.966(2) &   & 2 &  0.9875(08) & 0.998(2) \\
  & 3 &  0.9971(09) & 0.968(2) &   & 3 &  0.9977(08) & 1.000(2) \\
  & 4 &  1.0004(09) & 0.968(2) &   & 4 &  1.0020(10) & 0.998(2) \\
  & 5 &  1.0005(10)& 0.968(3) &   & 5 &  1.0032(11) & 0.997(2) \\
  & 6 &  1.0009(10)& 0.968(3) &   & 6 &  1.0018(10) & 0.997(2) \\
  & 7 &  1.0001(11)& 0.967(3) &   & 7 &  1.0016(10) & 0.997(3) \\
\bottomrule
\end{tabular}
\end{table}

As the data in Table~\ref{tab:expo} indicates, the optimization performs rather well. The first RG iteration generates an exponent of $y_t=1.0001(11)$ which is  within the statistical error of the exact value, and the second iteration generates $y_t=1.0016(10)$ which are close to the the exact value. The data consists of $10^6$ samples which is drawn independently from the training data set. It is surprising that the machine trained on such small training data of only $10^4$ examples is able to \textit{generalize} well and compute statistics based on a data set of  $10^6$ samples. Table~\ref{tab:expo} also contains values computed with majority-rule transformation for  comparison~\cite{swendsen1982}, giving $y_t=0.967(3)$ and $y_t=0.997(3)$ at the first and second RG iteration respectively.

\section{Conclusions}
We have parameterized our weight factor as a restricted Boltzmann machine to perform Monte Carlo renormalization group analysis. The optimal criterion for choosing the parameters is proposed to minimize the Kullback-Leibler divergence between the physical distributions and the normalized normalizing factors of the weight factors. It is shown that the set up is completely equivalent to learning with an restricted Boltzmann machine. 

Spin samples from the 2D Ising model produced by Monte Carlo were used to train the machine. Once trained, the machine is used to perform Monte Carlo renormalization group analysis and evaluate critical exponents. We show that the trained weight factor is optimal in that it faithfully reproduces the known exact thermal critical exponent within statistical error at the first step of renormalization transformation.

Our results demonstrate that the divergence minimization criterion may produce optimal convergence in the Monte Carlo renormalization group and may serve as a tool for more challenging problems such as the three-dimensional Ising model where the approach to the fixed point upon renormalization is known to be slow. Furthermore, our work may provide a statistical-mechanical point of view to the question of why deep learning works so well.

\section*{Acknowledgement}
This work was supported by Ministry of Science and Technology (MOST) of Taiwan under Grant numbers  108-2112-M-002 -020 -MY3,  and 107-2112-M-002 -016 -MY3.
 \bibliographystyle{apsrev4-1}
\bibliography{ref}

\end{document}